\documentclass[%
 sd,%
rsi,%
 amsmath,amssymb,
 reprint,%
]{revtex4-1}
\usepackage{eurosym}
\usepackage{color}
\usepackage{graphicx}
\usepackage{dcolumn}
\usepackage{bm}

\begin{document}


\title{High performance passive vibration isolation system for optical tables using six-degree-of-freedom viscous damping combined with steel springs}

\author{Gero L.\ Hermsdorf}
\author{Sven A.\ Szilagyi}
\altaffiliation[Current address: ]{Max-Planck-Institute for Solid State Research, Heisenbergstrasse 1, 70569 Stuttgart, Germany}
\author{Sebastian R\"{o}sch}
\author{Erik Sch\"{a}ffer}
\altaffiliation[Corresponding author. Email: ]{erik.schaeffer@uni-tuebingen.de}
\affiliation{Cellular Nanoscience (ZMBP), University of T\"{u}bingen, Auf der Morgenstelle 32, 72076 T\"{u}bingen, Germany.}
\date{\today}

\begin{abstract}
Mechanical vibrations in buildings are ubiquitous.  Such vibrations limit the performance of sensitive instruments used, for example, for high-precision manufacturing, nanofabrication, metrology, medical systems, or microscopy.  For improved precision, instruments and optical tables need to be isolated from mechanical vibrations.  However, common active or passive vibration isolation systems often perform poorly when low-frequency vibration isolation is required or are expensive.  Furthermore, a simple solution such as suspension from common bungee cords may require high ceilings.  Here we developed a vibration isolation system that uses steel springs to suspend an optical table from a common-height ceiling. The system was designed for a fundamental resonance frequency of 0.5\,Hz.  Resonances and vibrations were efficiently damped in all translational and rotational degrees of freedom of the optical table by spheres, which were mounted underneath the table and immersed in a highly viscous silicone oil.  Our low-cost, passive system outperformed several state-of-the-art passive and active systems in particular in the frequency range between 1--10\,Hz.  We attribute this performance to a minimal coupling between the degrees of freedom and the truly three dimensional viscous damping combined with a nonlinear hydrodynamic finite-size effect.  Furthermore, the system can be adapted to different loads, resonance frequencies, and dimensions.  In the long term, the excellent performance of the system will allow high-precision measurements for many different instruments.
\end{abstract}

\maketitle

\begin{figure*}
	\centering
			\includegraphics[width=\textwidth]{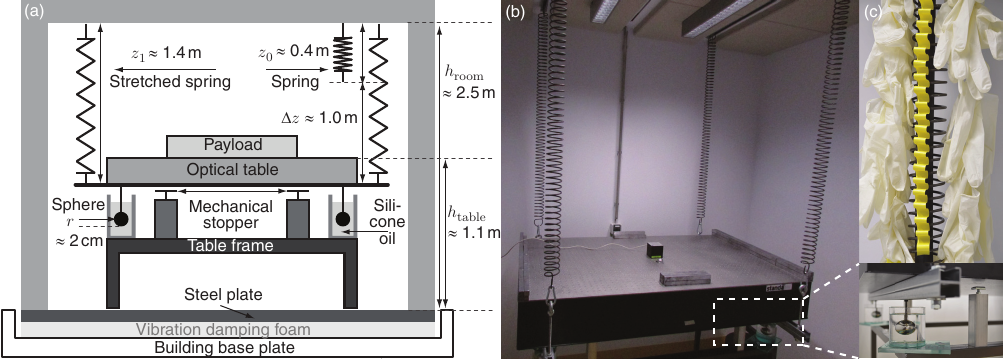}
	\caption{Steel spring based vibration isolation system.  (a)  Schematic of the system suspended from the ceiling of a walk-in chamber.  (b)  Picture of the system showing the vibration analyzer in the middle of the optical table and lead bricks as additional payload.  All vibration measurements were performed in this manner.  Magnified inset: View of the damping elements, mounting rail, and mechanical stopper.  (c) Damping implementation for the internal steel spring resonance based on rubber gloves and tape (yellow) [not shown in (b)].}
		\label{fig:Fig1}
\end{figure*}
\section{Introduction}
Advances in modern technology have enabled the investigation and development of nanoscale objects ranging from semiconductor devices to single molecular machines in biology.  One requirement to manipulate and observe such objects, is to isolate instruments that are used to characterize these objects from mechanical vibrations that are particularly present in fabrication plants and research laboratories.  The amplitude of such vibrations often exceeds the dimension of nanoscale objects in particular for frequencies below 10\,Hz.  The frequency band of building noise typically ranges from sub-1\,Hz to several tens of hertz depending on the source.  Typical noise  sources that couple to building resonances in the range of about 1--40\,Hz are elevators ($\lesssim$40\,Hz), people walking in the building ($\approx$1--5\,Hz), heating/ventilation/air conditioning ($\approx$7--350\,Hz), machines, motors, and transformers ($\gtrsim$4\,Hz, often with peaks close to the power line frequency of 50\,Hz or 60\,Hz and overtones), wind ($\approx$1--13\,Hz), ocean waves $\approx$\,0.1\,Hz and many more such as nearby traffic \cite{Ver2006,Harris2002}.  Vibrations with frequencies below $\lesssim$100\,Hz have a long wavelength compared to typical room dimensions.  Therefore, they are well transmitted through structures.  Because of their long wavelength they are also poorly damped by most materials.  Mechanical vibrations with frequencies above 100\,Hz often have sufficiently low amplitudes that they do not interfere with measurements.  In this frequency range, sound isolation is more important, which will not be considered further here.

For isolation of instruments from mechanical vibrations, most passive systems are approximated and oversimplified by a one-dimensional damped-harmonic-oscillator model \cite{nolting,lanlif,Ver2006,Harris2002,accurion,minusk1991}.  Below the resonance frequency $f_0$, vibrations directly couple to the instrument without attenuation.  Above resonance, vibrations are attenuated.  The relative amplitude of transmitted vibrations rolls off in analogy to a low pass filter.  The strength of the filter depends on the damping ratio $\zeta = \gamma / (2\sqrt{m\kappa})$, where $\gamma$ is the damping coefficient, $m$ the instrument mass, and $\kappa$ the spring constant.  For high frequencies $f\gg f_0$ the amplitude of a damped harmonic oscillator falls off with $1/f^2$ independent of the damping ratio.  However, the motion transmissibility of a vibration isolation system, taking into account the displacement of the oscillator position relative to the (moving) support, falls off with $1/f$ \cite{Ver2006,Harris2002}.  And, counterintuitively, the amount of vibration isolation for $f > \sqrt{2}f_0$ is less with an increased damping ratio.  Therefore, systems are typically not overdamped with the consequence that at resonance vibrations are slightly amplified \cite{Ver2006,Harris2002}.  Real systems have many degrees of freedom, typically coupled through viscoelastic damping elements \cite{Harris2002}. The response of such a system to a disturbance often cannot be solved analytically.  Thus, the vibration isolation performance needs to be tested under conditions that the system is designed for.  The optimal amount of damping will depend on the noise spectrum of the building and other criteria such as an optimized transient response.  Overall, for efficient vibration isolation between 1--10\,Hz, a passive vibration isolation system ideally should have a fundamental resonance frequency below 1\,Hz, be somewhat underdamped ($\zeta < 1$), and provide damping in all degrees of freedom without coupling them.

Typical vibration isolation systems for optical tables include bungee cords \cite{bungee}, air damped tables \cite{Voigtlaender2017}, passive systems with a negative-stiffness mechanism \cite{minusk1991}, active systems that include an accelerometer, an actuator and feedback controller \cite{richman08,kim12,accurion}, and less common pendulum systems \cite{steph91}.  While suspension from bungee cords is by far the cheapest solution, damping may not be optimal, cannot be tuned, and is determined by the choice of rubber.  Also, because rubber is viscoelastic there is creep in the extension, the stress-strain relation may be highly non-linear, and large static strains exceeding 50\.\% are not recommended over long periods of time \cite{Harris2002}.  One reason for the latter recommendation is that rubbers may crystallize under high, continuous strain causing failure of the material \cite{Harris2002}.  In general, for a ceiling-suspension system, the ceiling height limits the maximum length of the extended suspension spring given by its resting length plus its extension.

Low resonance frequencies require high ceilings.  Interestingly, the resonance frequency $f_0$ of a mass suspended from a ceiling via a Hookean spring only depends on the extension $\Delta z$ of the spring
\begin{equation}
    f_0 = \frac{1}{2\pi}\sqrt{\frac{\kappa}{m}} = \frac{1}{2\pi}\sqrt{\frac{g}{\Delta z}}\approx \frac{0.5\,\text{Hz}}{\sqrt{\Delta \tilde{z}}},
    \label{eq:resfreq}
\end{equation}
where the gravitational acceleration is denoted by $g$ and the value of the extension in meters by $\Delta \tilde{z}$.  For example, for a resonance frequency of 0.5\,Hz, 1\,m of spring extension is necessary.  The relation is a consequence of Newton's first law, that, in steady state, the table is at rest with no net force acting on it: the sum of the gravitational force $F_g$ and spring force $F_s$ is zero, i.e.\ $F_g + F_s = mg - \kappa \Delta z = 0$, resulting in Eq.~\ref{eq:resfreq} due to the linear relation between the spring constant and the mass ($\kappa = mg/\Delta z$).  Thus, for a given geometry, i.e.\ ceiling height, the resonance frequency is only limited by how far the suspension spring can be extended.  To achieve the maximum extension for a given mass, the spring constant needs to be chosen appropriately.  While rubber generally allows for very high strains up to several times the resting length, for continuous strain applications---as stated above---maximum strains of $\approx$50\,\% are recommended.  This strain also roughly corresponds to the maximum strain before failure of common bungee cords \cite{Martinez2014}.  Thus, with such a strain and an extension of one meter, the resting length of the bungee cord would be about two meters.  Together with an optical table height of about one meter, results in a ceiling height of about four meters.  Steel springs allow for a shorter resting length, are ideal for large static, continuous deflections, and practically have no creep when operated at room temperature \cite{Harris2002}.  However, steel springs require an additional damping system.  Here, we show how to implement a vibration isolation system that is viscously-damped in all degrees of freedom minimizing the coupling between them.  The system is based on steel springs and suitable for common ceiling heights below three meters.  Higher ceilings should allow for a lower resonance and even better performance.

\section{Materials and methods}
We used steel springs (Z209JX made of EN10270-1 steel; Gutekunst Federn, Metzingen, Germany) with a spring constant of $k=0.39$\,N/mm (initial tension of 21\,N and maximum spring force of $438\pm22$\,N) having an unloaded resting length of $z_0 = 0.365\pm0.004$\,m (including the mounting hooks), a maximum extension of 1.054\,m, and weight of $\approx$1.4\,kg.  The silicone oil had a very high viscosity of $\eta = 100$\,Pas (Wacker AK 100000, Wacker Chemie AG, Munich, Germany).  Note that the silicone oil was the most expensive component of the vibration isolation system.  The optical table had dimensions of 900$\times$1400$\times$200\,mm and a mass of around 146\,kg (1HT09-14-20; Standa, Vilnius, Lithuania).  As additional test payload, we used about 40 lead bricks weighing 1\,kg each.  The steel spheres (51604M6; ball-tech Kugeltechnik GmbH, Bodenheim, Germany) used as damping elements had a radius of $r = 2$\,cm, a mass of 245\,g, and have a M6 thread used to mount them below the optical table.  Transparent silicone oil beakers mounted on a table frame were made of acrylic glass in the local workshop with an inner diameter and height of 10\,cm.  To measure acceleration and velocity, we used a vibration analyzer system with a specified sensitivity in acceleration of 1\,$\mu g$ over a frequency range of 2--1000\,Hz (VA-2; The TableStable Ltd., Mettmenstetten, Switzerland).  Note that our designed resonance frequency of 0.5\,Hz was outside this range (the vibration analyzer's transfer function, approximated well by $1/(1+(0.73/f)^2)/(1+(f/858)^2)$ in the frequency range of 0.1--5000\,Hz, has a value of $\approx$0.3 at 0.5\,Hz) and vibrations on the table were largely below 1\,$\mu g$.  Also note that while vibration amplitudes may be underestimated outside the specified frequency range, the frequency itself of a ceratin vibration, e.g.\ a resonance, is still reliably measured.  From the acceleration amplitude $a$, the velocity and displacement spectral amplitudes can be calculated by $|\upsilon| = \frac{1}{\omega}|a|$ and $|x| = \frac{1}{\omega^2}|a|$, respectively, where $\omega = 2\pi f$ is the angular frequency.  To compare the acceleration on the optical table to the gravitational acceleration, we plot reference lines with a constant spectral density of acceleration in units of $g$.  For example, the root-mean-square (rms) velocity density corresponding to 10\,n$g$ is calculated according to $1/(2\pi f)\cdot9.81\cdot10^{-8}/\sqrt{2}$\,m/s$^2$/$\sqrt{\text{Hz}}$.  To record the power spectral densities of the vibrations, we used a data acquisition system from National Instruments operated via LabView with custom written software.  For noise reduction, we always averaged 40 power spectra.  To measure a macroscopic transient response (1--2\,cm displacement from the equilibrium position) of the vibration isolation system, we mounted a laser (LuxX 488-100, Omicron-Laserage Laserprodukte GmbH, Rodgau-Dudenhofen, Germany) on the optical table pointing at a camera (PowerShot SX500IS) fixed to the inside wall of the walk-in chamber.  The recorded video was analyzed in Fiji \cite{Schindelin2012} by tracking the position of the laser spot as a function of time.

\begin{figure*}
	\centering
	\includegraphics[width=\textwidth]{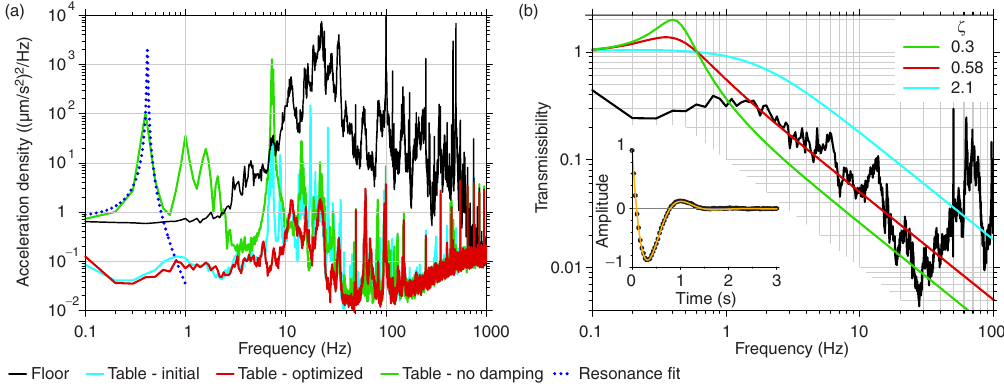}
	\caption{Vibration spectrum, transmissibility and transient response of the vibration isolation system. (a) Power spectral density of the acceleration measured on the floor and optical table. (b) Transmissibility of the optimized table as a function of frequency.  The transmissibility $T = ([1 + (2\zeta f/f_0)^2]/[(1-[f/f_0]^2)^2 + (2\zeta f/f_0)^2])^{\frac{1}{2}}$ was fitted to the data (red line) in the range of 1--40\,Hz using a fixed value of $f_0 = 0.42$\,Hz.  With the same $f_0$, $T$ for $\zeta = 0.3$ and 2.1 are shown for comparison.  Inset: Normalized amplitude (data: circles, fit: yellow line) after a $\approx$1\,cm displacement from the equilibrium position as a function of time.  As a guide to the eye, the edge of the inset has a $1/f$ slope.}
	\label{fig:Fig2}
\end{figure*}
\section{Instrument Design}
We designed the vibration isolation system for an optical table with a resonance frequency of $\approx$0.5\,Hz in a walk-in chamber with a ceiling height of $h_{\text{room}} \approx 2.5$\,m [Fig.~\ref{fig:Fig1}(a)].  The chamber is located in a basement laboratory room of the building.  This location already reduces the input of vibrations significantly.  The walk-in chamber itself is isolated from the remaining building via a vibration damping foam underneath a steel plate floor.  On the steel plate, the chamber is made of brick walls with a concrete ceiling and sound proof door.  The chamber isolates well from acoustic noise.  Experiments are controlled from outside the chamber minimizing user-induced disturbances.  The walk-in chamber is equipped with mounting rails in the ceiling.  From these ceiling rails, we suspended an optical table using steel springs.  The springs were attached to the table via steel mounting rails on which the optical table was fixed [Fig.~\ref{fig:Fig1}(b)].  Note that we used small rubber pads between the end of the springs and the rails to reduce the transmission of high-frequency vibrations.  To the same rails, we attached the damping elements---steel spheres immersed in silicone oil.  An additional table frame under the optical table was used to fix the silicone oil containers.  To prevent extreme downward displacements, we mounted mechanical stoppers to this table frame [see magnified inset in Fig.~\ref{fig:Fig1}(b)].  Based on the estimated weight, the spring constant was chosen to achieve $f_0 \approx 0.5$\,Hz.  With the total suspended mass of the optical table and test payload (eventually a high-resolution microscope), springs were extended to their maximum with a total length of $z_1 = 1.420$\,m resulting in $\Delta z = z_1 - z_0 = 1.055$\,m.  This extension corresponds to a mass of $m \approx 190$\,kg and resulted in a table height of $h_{\text{table}} \approx 1.10\,$m.  An internal resonance of the steel springs, their surge frequency, was effectively damped with soft rubber contacts of standard latex laboratory gloves hanging from the springs and tape connecting the spring coils [Fig.~\ref{fig:Fig1}(c)].  As an equally-well-performing alternative, we used long stripes cut from an inner tube of a bicycle tire mounted on the inside of the springs.  With this additional damping, the vibration isolation system is complete for characterization and performance measurements.

\section{Results}
To characterize the performance of the vibration isolation system, we measured the power spectral density (PSD) of the vibrations [Fig.~\ref{fig:Fig2}(a)].  We placed the sensor of the vibration analyzer system either in the center of the optical table [Fig.~\ref{fig:Fig1}(b)] or on the floor directly beneath the optical table and recorded the vibrations from the control room outside the walk-in chamber.  The PSD of the floor acceleration inside the walk-in chamber [black line in Fig.~\ref{fig:Fig2}(a)] had a maximum at $\approx$20\,Hz with a first peak at $\approx$11\,Hz presumably corresponding to its lowest fundamental resonance.  A narrow peak at 100\,Hz was caused by the air-conditioning outside the walk-in chamber, which was significantly reduced when the air-conditioning was turned off.  On the optical table, vibrations were significantly reduced at all frequencies [cyan line in Fig.~\ref{fig:Fig2}(a)].  With the initial implementation of the system, we observed high peaks at around 7\,Hz, 9\,Hz, and corresponding overtones. These peaks originated from an internal resonance or surge frequencies between the individual coils of the steel springs due to their finite mass.   We successfully damped these resonances by weak rubber contacts between the coils on both the in- and outside of the springs [Fig.~\ref{fig:Fig1}(c)].  These loose contacts optimized the damping in the frequency band around 1--10\,Hz [red line in Fig.~\ref{fig:Fig2}(a)].  Note that above 40\,Hz, the measurement was largely limited by the sensitivity of the vibration analyzer.  Also, some of the peaks may correspond to electronic noise (overtones of the power line frequency).   When we removed the damping elements below the optical table, the PSD showed a resonance peak [green line in Fig.~\ref{fig:Fig2}(a)] fitted to be at $f_0 = 0.42\pm0.01$\,Hz (blue dotted line) roughly consistent with the expected value.  Thus, the table was well isolated from typical laboratory vibrations with frequencies $f \gtrsim 1$\,Hz.

To quantitatively assess the table performance, we estimated its transmissibility and measured its transient step-response behavior.  Since we could not measure vibrations on the ceiling or on top of the walk-in chamber, we approximated the motion transmissibility by the ratio of the vibration amplitude on the table relative to the floor [Fig.~\ref{fig:Fig2}(b)].  The measurement was limited to a frequency range of $\approx$2--40\,Hz because the vibration analyzer could not reliably measure vibration amplitudes at lower frequencies and, as mentioned above, was not sensitive enough at higher frequencies.  In the reliable range, the transmissibility decreased with the expected $1/f$ dependence. A best fit of the theory, resulted in a damping ratio of $\zeta = 0.58\pm0.01$.  The lowest measured transmissibility was about 0.005 corresponding to -45\,dB.  Thus, based on the transmissibility, as designed, the vibration isolation system performed as a slightly underdamped ($\zeta \lesssim 1$) harmonic oscillator.  After a step-like disturbance, such a system should exponentially relax back to its equilibrium position with some ringing oscillations.  This transient behavior, we indeed observed [inset Fig.~\ref{fig:Fig2}(b)].  A fit of an exponentially damped oscillation resulted in a time constant of the exponential relaxation of $\tau = 0.36\pm0.01$\,s and an oscillation period of $T = 1.39\pm0.01$\,s.  Thus, with a sub-second relaxation time constant, macroscopic disturbances were quickly damped.

\begin{figure*}
	\centering
	\includegraphics[width=\textwidth]{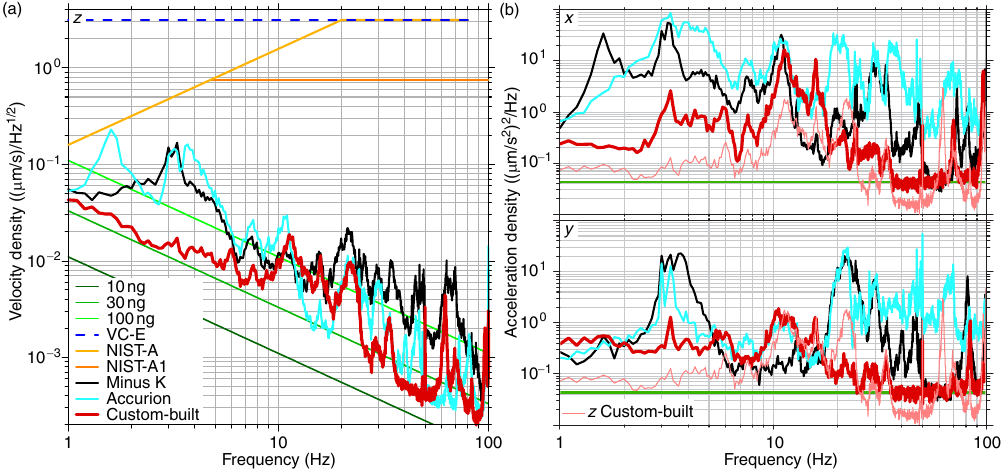}
	\caption{Performance comparison.  (a) Vibration rms velocity density for the vertical direction on top of optical tables isolated using a Minus K 500BM-1 (Minus K Technology, Inglewood, USA), Halcyonics\_VarioBasic\_90-300 (Accurion, G\"{o}ttingen, Germany), and our optimized custom-built system.  Also shown are vibration criteria (VC-E, NIST-A \& A1) for facilities with sensitive equipment and reference lines in units of the gravitational acceleration. (b) Horizontal acceleration power spectral density for the same systems using the same color scheme as in (a).  For comparison, the data for the vertical direction of our system [red line in Fig.~\ref{fig:Fig2}(a)] is shown as a thin pink line.}
	\label{fig:Fig3}
\end{figure*}
To evaluate the overall performance of our custom-built system, we characterized its performance relative to two commercial systems and common vibration criteria (Fig.~\ref{fig:Fig3}).  Since vibration criteria are expressed in terms of rms velocity (the square root of the vibration velocity PSD; see Materials and Methods for the conversion of the measured acceleration PSD to a velocity PSD), we measured the rms velocity in the vertical $z$-direction on three similar optical tables with similar weights (including the payload) in the same laboratory room each placed in similar walk-in chambers having comparable floor vibration amplitudes [Fig.~\ref{fig:Fig3}(a)].  Two tables were isolated from vibrations from a state-of-the art active and passive system, respectively, and one by our optimized custom-built system.  While our system performed best between 1--10\,Hz, above 10\,Hz all three systems performed similar.  Below 1\,Hz, we could not measure any difference between the devices limited by the vibration analyzer.  All systems performed better than the stringent NIST-A1 norm and well below the VC-D vibration criterion, which is 2-fold higher than VC-E and the recommended standard for SEM and TEM electron beam devices.  The vibration level on the optical table isolated with our custom-built system was comparable to that induced by an acceleration of about 30\,n$g$ for most frequencies.  Since viscous damping should perform equally well in all directions, we also compared the vibrations in the two horizontal directions [Fig.~\ref{fig:Fig3}(b)].  For both horizontal degrees of freedom, with few exceptions in narrow frequency bands, our system outperformed the commercial systems significantly in the 1--100\,Hz range.  The good performance in all spatial directions indicates little coupling between the different translational degrees of freedom.

\section{Discussion}
Our custom-built vibration isolation system was designed as a slightly underdamped harmonic oscillator with a resonance frequency of about 0.5\,Hz.  Since the springs were extended to their maximum extension, the total suspended weight was $m\approx187$\,kg (maximum load plus initial tension divided by gravitational acceleration) consistent with our weight estimate.  Using this value, the measured resonance frequency of 0.42\,Hz, and Eq.~\ref{eq:resfreq}, the spring constant of the individual springs was 0.33\,N/mm---somewhat smaller than the specifications.  Based on the measured extension and Eq.~\ref{eq:resfreq}, the resonance frequency should have been 0.49\,Hz.  The frequency of 0.42\,Hz corresponds to an extension of 1.42\,m according to Eq.~\ref{eq:resfreq}.  Interestingly, this value corresponds exactly to the springs resting length plus its extension, $z_1 = z_0 + \Delta z$, implying that this length was the decisive length.  Overall, we attribute the differences from the expected values to nonlinearities of the maximally extended springs deviating from the Hookean approximation.  Apart from this small nonlinear response, we do not expect any other effects, like creep or failure, to occur when operating the springs at their maximum specified extension.

The oscillation period of the transient response was shorter compared to the expected period of the fundamental, ``bounce'' or ``heaving'' mode of the optical table.  Thus, the table relaxed via a different mode.  Since the springs and/or damping elements are not identical, each corner of the table relaxed with a different time constant resulting in a rotation around the center of mass of the table.  The resonance frequency of this rotation for a weakly or uncoupled system is approximately $f_{\text{rot}} \approx f_0 \sqrt{3(1-4\frac{\Delta L}{L})}$, where $L$ is the length or width of the table and $\Delta L$ the distance from the edge of the table to the point of suspension using $\frac{1}{12}mL^2$ for the table's moment of inertia \cite{Ver2006,Karioris1992}.  With the values $L = 1.4$\,m and $\Delta L = 0.12$\,m, the period for the rotational oscillation is $T_{\text{rot}} = 1.4$\,s, which is in excellent agreement with our measured period.  Thus, the table did not relax via its heaving mode, but rather by a rocking, rolling or pitching mode around its center of mass.  Since the resonance frequency of this mode is higher compared to the fundamental frequency, the transient response was faster.  Furthermore, the quantitative agreement of the measured rotational oscillation period with the theory implies that the coupling between translational and rotational degrees of freedom was small \cite{Karioris1992}.

The transient response provides information about the amount of damping.  Based on the measured mass and exponential relaxation time of the transient response $\tau = 2m/\gamma$, the systems's damping coefficient was $\gamma = 1.04$\,kNs/m.  This value is 6.9$\times$ larger compared to the Stokes drag of the spheres of $\gamma_0 = 4 \cdot \,6 \pi \eta r \approx 0.15$\,kNs/m, where $\eta$ is the viscosity of the silicone oil and the factor 4 accounts for the four damping elements.  The difference can be explained by the nearby walls of the oil container.  The drag of a sphere along the axis of an infinite cylinder with a distance to the cylinder wall, in our case, corresponding to 2.5 times the sphere's radius is increased by a factor of $\approx$3.5 compared to Stokes drag (e.g.\ \cite{HappelBrenner}, p.\ 318).  In addition, the finite length of the cylinder needs to be accounted for, whereby our dimensions invalidate a linear approximation \cite{Sano1987}.  A lower estimate of the drag increase is given by how the drag increases for movements perpendicular to an infinite flat wall.  In our case, this increase is $\approx1.8\gamma_0$ \cite{Schaeffer2007}.  Multiplying these two factors results in a total increase of about $6.3\gamma_0$, close to our measured value.  Thus, for macroscopic displacements, the damping coefficient is significantly increased by the finite size and geometry of the oil container.

While the measured resonance and mass are consistent with our expectations, there is a discrepancy with respect to the measured damping coefficient and the expected transmissibility.  Using the measured values for the damping coefficient, mass and spring constant results in a damping ratio of $\zeta \approx 2.1$.  With $\zeta > 1$, the system should be overdamped inconsistent with the damped oscillatory transient response.  We attribute the oscillation to the relaxation via the rocking mode and, possibly, finite-size flow effects of the viscous oil in the cylinders.  Also, based on this damping ratio, theoretically the transmissibility should be worse [cyan line in Fig.~\ref{fig:Fig2}(b)].  For comparison, we also plotted the transmissibility for $\zeta = 0.30$ (green line)---the expected value without the wall effect.  The best-fit resulted in $\zeta \approx 0.58$, a value that is closer to the Stokes drag estimate without walls.  One possible explanation might be a non-linear, time-dependent viscous response.  In the absence of large disturbances, the vibration amplitude on the table was $\approx$10\,nm at 1\,Hz falling off with roughly $1/f^2$ (i.e.\ 0.1\,nm at 10\,Hz).  These amplitudes are much smaller than the dimensions of the spheres and cylinder used for damping.  If the spheres move with these amplitudes on these time scales relative to the stationary cylinder, the full equilibrium flow profile in the cylinders may not have been established \cite{Felderhof2009} resulting in an effective damping coefficient closer to the Stokes drag estimate.  For small, short, and random amplitude fluctuations, the spheres effectively may not ``feel'' the presence of the walls.  A non-linear damping coefficient that increases with deflection amplitudes could also explain the transient ringing behavior.  Alternatively, since the walk-in-chamber is a vibration isolation system in itself, we have a multistage system that improves high frequency attenuation.  For frequencies well above the table and chamber resonances, the transmissibility of the combined system should roll off with 1/$f^2$ \cite{Ver2006,richman08}.  However, since the resonance of the chamber is about 11\,Hz, this effect should only occur for significantly larger frequencies.  At these frequencies, the measured transmissibility showed a broad peak at around 70\,Hz.  We attribute this peak to resonances of the table frame that supports the oil containers.  A more rigid construction of the latter should improve the performance further.  However, since at these frequencies the displacement amplitudes of vibrations are already on the sub-{\AA}-level, we did not pursue this idea further.

Overall, our vibration isolation system combined the advantages of steel springs with viscous damping.  Steel springs do not drift or creep and allow for a maximum extension in rooms with a common ceiling height allowing for good low-frequency isolation.  Since springs are available in all dimensions, our design can be adjusted to different payloads and ceiling heights.  For example, a 4-m high ceiling should allow for a resonance of $\approx$0.3\,Hz.  We could reduce the high-frequency transmission through the springs and internal resonances by adding soft damping elements to the springs themselves.  The rocking motion inherent to the system was beneficial in the sense that it reduced the transient response time.  The viscous damping based on four spheres, has the advantage that all translational and rotational degrees of freedom are damped simultaneously.  For the horizontal translational modes of freedom, the resonance frequency is given by the well-known pendulum resonance.  Since the pendulum length here roughly corresponds to the spring resting length plus its extension, $z_1 = z_0 + \Delta z$, the horizontal resonance is given by $f_0^{\rm{horiz}} \approx \frac{1}{2\pi}\sqrt{\frac{g}{z_1}}$, which is even lower compared to the vertical direction (Eq.~\ref{eq:resfreq}).  Thus, horizontal vibration isolation is expected to be at least as good as for the vertical direction.  This expectation is supported by our data [Fig.~\ref{fig:Fig3}(b)].  Since the optical table's rotational degrees of freedom correspond to linear combinations of the translational degrees of freedom of the four individual damping elements, a good performance for the translational degrees of freedom also implies good performance for the rotational degrees of freedom with small coupling between the individual degrees of freedom as pointed out above.  Therefore, the viscous elements provide independent damping for each of the three fundamental translational and rotational degrees of freedom of the optical table.

The amount of viscous damping can be adjusted by the size of the spheres, the viscosity of the oil, and the distance of the spheres to the bottom of the oil containers.  Since the drag coefficient diverges as the spheres approach the bottom \cite{Schaeffer2007}, the damping coefficient can roughly be varied 10-fold using the latter approach.  A lower damping ratio compared to the one we used, may reduce the transient response time, but will increase low frequency noise.  The optimal damping depends on the application and vibrational noise spectrum.  While commercial systems are very compact and are designed to fit under an optical table, our system requires ceiling mounting and space for the springs, which may limit some applications.  The better performance of our system compared to the commercial ones may be due to the truly viscous damping, which provides damping in all six degrees of freedom and may minimize the coupling between these degrees in contrast to common viscoelastic dampers.  Also, our higher damping ratio may more efficiently reduce the ringing amplitudes of transients arising from the random, superimposed step-like disturbances coming from the building.  Overall, the performance of our system meets stringent vibration criteria---it is better than VC-K---with a vibration level of about 30\,n$g$ in the vertical direction.  The system is comparable to, and in the low frequency range better than, designed low-vibration laboratories \cite{Iwaya2011,Voigtlaender2017}.  Our solution is cheap, simple to build, and possible to be scaled for different payloads.  Thus, in the long term, we expect that our custom-built, high performance vibration isolation system can be used for many other delicate measurement devices such as superresolution or electron microscopes and will enable sensitive experiments by effectively isolating the instruments from vibrations.

\section*{Author Contributions}
E.S., G.L.H., and S.A.S.\ designed the research, G.L.H., S.A.S., and S.R.\ built the system, G.L.H., and S.A.S.\ performed measurements, G.L.H., S.A.S., and E.S.\ analyzed the data, and G.L.H., and E.S.\ wrote the manuscript.

\begin{acknowledgments}
We thank Steve Simmert, Anita Jannasch, Mayank Chugh, and Michael Bugiel for comments on the manuscript.  G.L.H.\ acknowledges financial support from the International Max Planck Research Schools from Molecules to Organisms, Max Planck Institute for Developmental Biology, T\"{u}bingen.    This work has been supported by the Deutsche Forschungsgemeinschaft (DFG, CRC1011, Project No.\ A04), the European Research Council (ERC-POC Project No.\ 755161, PRIMASKOTI), and the University of T\"{u}bingen.
\end{acknowledgments}

\providecommand{\noopsort}[1]{}\providecommand{\singleletter}[1]{#1}%

\end{document}